# Electron microscopy and diffraction study of high-temperature diamond-like Si-based ferromagnetic with self-organized super-lattice distribution of Mn impurity

E. S. Demidov, E. D. Pavlova, A. I. Bobrov, V. V. Karzanov, N. V. Malekhonova, A. A. Tronov

*Nizhni Novgorod state university, pr. Gagarina 23, Nizhni Novgorod, 603950 Russia*
demidov@phys.unn.ru

New structure data for the diluted magnetic semiconductor (DMS) Si:Mn with Curie point ≈500K, synthesized by laser method are presented. The higher resolution transmission electron microscopy and diffraction was applied in directions <110> and <100> of epitaxial layers of DMS Si:15%Mn with elimination of the contribution from GaAs substrate and interface. It is established, that the DMS Si:Mn can be represent as compound with variable composition $Si_{2+x}Mn_{1-x}$ ($0 < x < 1$), single-phase diamond like structure, high crystal perfection and the self-organized formation of the super-lattice structures with the period equal to triple the distance between the nearest (110) atomic layers and interval between (110) layers which are doped by Mn atoms and orientated along the direction of Si:Mn film growth. The Si:15%Mn (or $Si_{2.5}Mn_{0.5}$) films consist of blocks with the 15-50-nm-sizes and with mutually perpendicular orientations of the super-lattice modulations. Atoms of manganese in the (110) layers, doped by these impurity, settle down in the form of strips of an one-atom thickness, width from one to several nm, with a length stretched within all block. These manganese strips in the given (110) layer alternate with silicon strips of one-atom thickness and occupy about half of area of a layer (110) in the full consent with 15 % Mn according the spectral X-ray analysis and ferromagnetic resonance data.

The first ferromagnetic so-called diluted magnetic semiconductors (DMSs) based on Mn doped diamond like II-VI [1,2] and III-V compounds such as $Ga_{1-x}Mn_xAs$ and $In_{1-x}Mn_xAs$ with the atomic content of manganese impurity $x≈0.05$ [3] had been synthesized for the first time in the 1990s. To now numerous works concerned $Ga_{1-x}Mn_xAs$ DMS epitaxial layers with the perfect crystal structure. However, the maximum Curie point for this ferromagnet turns out to be below room temperature and does not exceed 170K [4]. The silicon_based DMSs are of special interest for applications owing to the predominant usage of silicon in the present-day technologies. There were many attempts to synthesize silicon-based ferromagnetic DMSs with 3$d$ dopants. The synthesis of such DMSs is even more complicated task than that in the case of III–V semiconductors. The near-equilibrium solubility (about $10^{16}$ cm$^{-3}$) of the dopants of the iron group from the middle of the 3$d$ series of the periodic table in Si turns out to be two orders of magnitude lower than that in the case of III–V compounds. These dopants occupy mainly interstitial sites and play the role of donors [5, 6]. The first Si:Mn/Si single-crystalline layers with 5% Mn exhibiting ferromagnetism up to 400 K (according to the magnetization data) were produced by electron-beam deposition [7, 8]. However, they have a relatively high resistivity (0.25–2.5 Ω cm) and are characterized by a semiconductor behavior, $ρ = ρ_0 \exp(E_a/kT)$. This means that their charge carrier density is insufficient for attaining the state of a degenerate semiconductor. This state is needed for the existence of the Fermi surface and for the physical implementation of the Ruderman–Kittel–Kasuya–Yosida (RKKY) indirect exchange interaction via the hole charge carriers, which is currently considered as the most probable mechanism of the ferromagnetic spin ordering of the magnetic 3$d$ ions in DMSs [9, 10]. The absence of degeneracy, i.e., the charge carrier density in silicon below $10^{20}$ cm$^{-3}$, suggests that the ferromagnetism reported in [7, 8] results from some inclusions of the second phase. The first electronically perfect (without inclusions of the second phase) sin_gle_crystalline silicon layers with the Mn dopant content ranging from 2 to 5% were formed by the implantation of Mn$^+$ ions with an energy of 200 keV at the irradiation doses of $(0.5–5) × 10^{16}$ cm$^{-2}$ and with subsequent annealing at 600–900 K for 5 min in a nitrogen atmosphere [11]. According to the magnetization data, their Curie temperature was not high (about 70 K). Hole conductivity was observed, but the hole density (up to $1.2 × 10^{18}$ cm$^{-3}$) was not sufficient for attaining degeneracy. Probably, the ferromagnetism here is due to the inhomogeneous distribution of manganese and to the existence of regions with the peak density of this dopant exceeding $10^{20}$ cm$^{-3}$. A similar implantation of Mn$^+$ in Si allowed the authors of [12] to form Si:Mn DMS layers which exhibited ferromagnetism up to 305 K according to the data on the magnetization and the Faraday effect within the 1–6 μm wavelength range.

The most high-temperature DMSs based on diamond-like semiconductors were synthesized in our laboratory by pulsed laser-plasma deposition [9, 13, 14]. We demonstrated the potentialities of such highly nonequilibrium technology in the synthesis of thin (30–200 nm) GaSb:Mn and InSb:Mn layers with the Curie temperature $T_C$ exceeding 500 K and Ge:Mn, Si:Mn, and Si:Fe layers with $T_C$ up to 400, 500, and 250 K, respectively, on GaAs, Si, and sapphire (Al$_2$O$_3$) single-crystalline substrates. The Si:Mn DMSs with 10–15% Mn, which were the most actively studied by us, are characterized by the highest mobility of charge carriers. In these layers, Mn atoms exhibit almost completely their electrical and magnetic activity. Such DMSs have a certain regular structure having no relation to ferromagnetic inclusions. This conclusion was indirectly verified by experiments where ferromagnetism was destroyed by heating or ionic irradiation and by comparison with the results obtained for digital alloys including the same components. This structure cannot be determined either by the usual X-ray diffraction method owing to the insufficient thickness of the layers and to small atomic mass of silicon or by the reflection electron diffraction technique because of the amorphous near-surface oxide layer masking the results concerning the film itself. In the previous work [15] the crystal structure of nanofilms of a Si:Mn DMS with the Curie temperature of about 500 K obtained by the pulsed deposition from a laser plasma has been studied by high-resolution transmission electron microscopy (HRTEM) and local electron diffraction (LED). The direct evidences demonstrating the regular structure in Si:Mn/GaAS DMS were presented. It is shown that the nonequilibrium laser technique allows achieving a pronounced supersaturation of the Mn solid solution as high as 15%. In such solutions, Mn substitutes silicon, leaving unchanged the diamond_like crystal structure and does not prevent the epitaxial growth of Si:Mn films. At the same time, there occurs a self_organized formation of the superlattice structure with the period equal to triple the distance between the nearest (110) atomic layers, where (110) layers doped with Mn are oriented along the growth direction of the Si:Mn film.

The laser-plasma deposition of Si:Mn DMS layers onto single-crystalline GaAs substrates with the (100) orientation is described in [9, 13, 14]. The HRTEM and LED measurements for the cross cut of Si:Mn DMS layers were performed using a JEOL JEM-2100F transmission electron microscope. We studied 50-nm-thick Si:Mn layers produced at 300°C, where, similarly to [13], the atomic content of Mn was equal to 15% (according to the data of the X-ray diffraction analysis with electron excitation). We observed the magneto-optic Kerr effect and the anomalous Hall effect, as well as high hole conductivity corresponding to the resistivity $\rho = 2.5 \times 10^{-4}$ $\Omega$ cm and high hole mobility $\mu = 33$ cm$^2$/V s. The hole density $p \geq 7.5 \times 10^{20}$ cm$^{-3}$ corresponds to the clearly pronounced degenerate semimetallic state of silicon, which means that no less than 10% of manganese dopant atoms are electrically active. The room-temperature ferromagnetic resonance (FMR) spectrum, similarly to that reported in [9, 13, 14], consists of several resonance absorption peaks. On cooling, these peaks merged, forming a single relatively narrow peak. Using the commonly assumed value 5/2 for the spin of manganese, we can deduce from the FMR data at 93 K the value on the manganese content, $N_{Mn} = 8 \times 10^{21}$ cm$^{-3}$. This value is nearly the same as that determined from the X-ray spectral analysis $N_{Mn} = 7.5 \times 10^{21}$ cm$^{-3}$, or 15 at % Mn. In other words, all manganese atoms are magnetically active. Note that silicon layers deposited using the same technique but without the manganese doping were paramagnetic and exhibited the electron type of conductivity with an electron density of about $10^{16}$ cm–3, which is five orders of magnitude lower than the charge carrier density in Si:Mn layers. Thus, the ferromagnetism and semimetallic hole conductivity of our Si:Mn DMS are due to the manganese dopant.

In [15] the HRTEM and LED of DMS Si:Mn layers cross-section were performed in one direction <110>. Thus details of distribution Mn in the (110) layers doped by these impurity in the super-lattice structure remain not clear - in the form of a single-atom thickness threads or a flat inclusions, what is picture of distribution Mn in other directions of <110> type, what nature of the Si:Mn layer nonuniformity, visible in **Fig. 1** in [15]. In the present work all these details made clear by the same HRTEM and LED as for cross-section cut in the <110> direction, and in the <100> direction of the same Si:Mn 50-nm-layers on GaAs(100), thinned in the layer plane. For this purpose, as it is schematically shown **in Fig. 1**, around the place of 300-nm-diameter electron beam path in perpendicular orientations of super-lattice planes are shown. On the right the crystallographic directions of the super-lattice modulations in various blocks are designated.

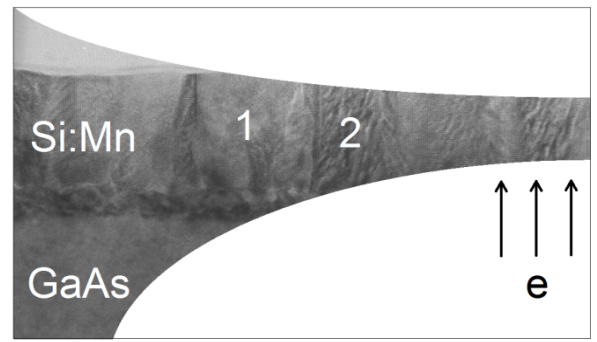

**Fig. 1**. The schematical view of part of cross-section cut of the sample for High-resolution transmission electron microscopy and Local electron diffraction in <100> direction of 50-nm-thick Si:15%Mn layer deposited by the laser-plasma technique at 300°C onto the single-crystalline GaAs substrate with the (100) orientation. The HRTEM lattice image of the cross-section cut of the same layer along a <110> crystallographic direction is used. By arrows the direction and place of electrons beam at carrying out HRTEM and LED is schematically shown perpendicularly plane of Si:Mn layer along the direction <100>.

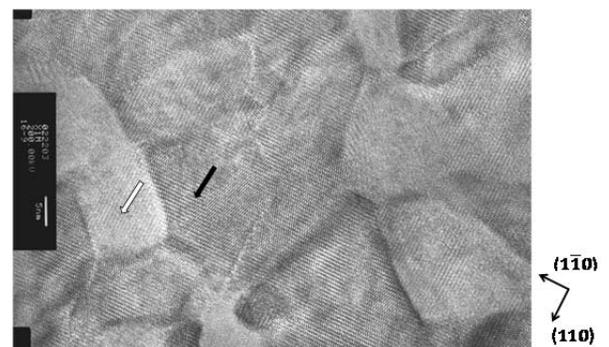

**Fig. 2**. High-resolution transmission electron microscopy plane-view image of the 50-nm-thick Si:15%Mn layer deposited by the laser-plasma technique at 300°C onto the single-crystalline GaAs substrate with the (100) orientation. The image is obtained along the ⟨100⟩ lattice direction. By white and black arrows two neighbor blocks with mutually

JEM-2100F device the GaAs substrate and disordered interface layer between this substrate and layer Si:Mn was removed by precision spherical mechanical polishing and the subsequent sliding ionic sputtering, the amorphous oxidized layer on other side of Si:Mn layer was removed by ionic sputtering.

In **Fig. 2**, we show the HRTEM image along the <100> lattice direction of the 50-nm-thick Si:Mn layer deposited by the laser-plasma technique at 300°C onto the single-crystalline GaAs substrate with the (100) orientation. Apparently, in the consent with data in [15], the Si:Mn film in the bulk consists of areas with super-lattice crystal modulation of structure in directions of <100> type with the period equal to the tripled distance between (110) type planes. "Equality" of perpendicular directions of modulation $<110>$ and $<1\bar{1}0>$ is represented that Si:Mn film consists of blocks with the 15-50-nm-sizes and with mutually perpendicular orientation of super-lattices. This is more clearly visible in fig. 3 where small areas of images of these blocks are shown with higher resolution.

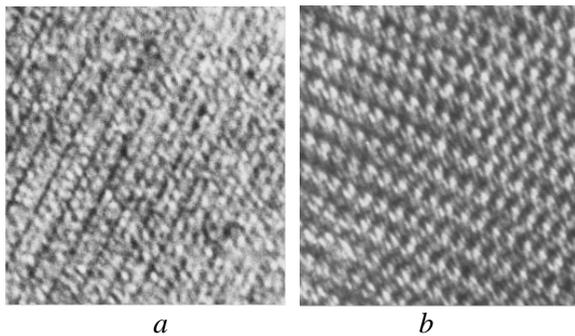

*a*         *b*

**Fig. 3.** The enlarged images shown by arrows in **Fig. 2** of small areas of the image of Si:15%Mn DMS layer, *a* –the area in fig. 2 of block marked by white arrow, *b* - of block marked by black arrow.

Conclusions from the HRTEM data in **Fig. 2** and **Fig.3** are confirmed by results of LED in <100> direction for DMS Si:Mn layer in **Fig. 4** where the similar diffraction pattern of GaAs single-crystal sample is resulted for comparison. On diffraction Si:Mn image there are the same reflections and same well-defined, as in the case of GaAs. It means, that investigated

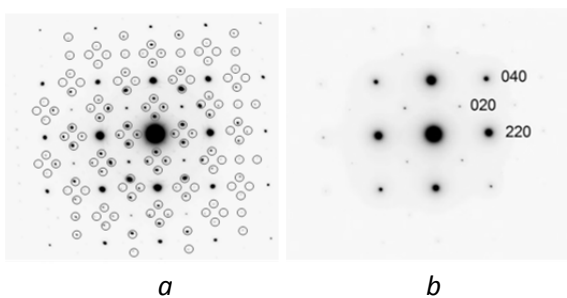

*a*         *b*

**Fig. 4.** Local electron diffraction patterns along the <100> direction of *a*- Si:Mn DMS layer and *b* – of single-crystal GaAs. Circles indicate the additional to electronic diffraction GaAs reflections.

Si:Mn DMS has diamond-like structure and high crystal perfection. Besides on the image in inverse space of Si:Mn there are the additional reflections noted on **Fig. 4a** by grey circles. Three times smaller, than at the basic reflections the distance between the neighbor additional reflections along <110> directions confirms the established in [15] self-organized formation of super-lattice with the period equal to the triple distance between the nearest (110) atomic layers and interval between (110) layers which are doped by Mn atoms and which are oriented along growth direction of Si:Mn film. It is obvious, that the picture in **Fig. 4a** is superposition of diffractions from both types of blocks with mutually perpendicular each other orientations of super-lattice modulations along both <110> type directions in the Si:Mn film plane. Diameter of the 300 nm electronic beam in several times exceeds the sizes of blocks of 15-50 nm and locks on blocks of both types. Absence of the diffraction reflections on pattern of cross-section cut of Si:Mn layer, shown in [15], means, that in all blocks the (110) planes enriched by manganese are oriented perpendicularly substrate plane in the direction of growth of this layer.

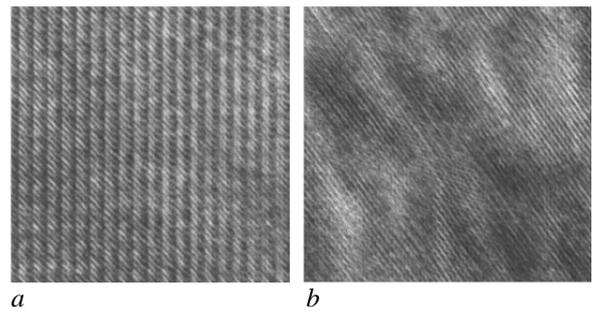

*a*         *b*

**Fig. 5.** The enlarged images shown by white figures on **Fig. 1** of small sites of the image of Si:15%Mn DMS layer along the <110> direction; *a* - the site 1 in **Fig. 1**, *b* - the site 2.

Presence of the established block structure of our Si:Mn laser layers allows us to understand the reason of presence of two kinds of areas on the image of a cross-section cut along the <110> direction, designated in white figures 1 and 2 in **Fig. 1**. Small sites of these areas with higher resolution are shown in **Fig. 5**. It is clear that areas of the kind 1 correspond to the look-through image of Si:Mn crystal along the planes of super-lattice modulation. Areas of the kind 2 correspond to the image in the direction normal to planes of super-lattice modulation. The variation of contrast within areas of the kind 2 apparently is caused by spatial change of Mn concentration in the monoatomic (110) layers doped by these impurity. It means, that DMS Si:Mn can be considered as compound of variable composition $Si_{2+x}Mn_{1-x}$ with the limiting chemical stoichiometric formula of alloy $Si_2Mn_1$ which corresponds to the maximum atomic share of manganese about 33 %. And this compound possibly keeps diamond-like structure at change $x$ from 0 to 1. Essentially important result here and in [15] consists in an establishment of possibility of formation of single-phase compound of variable composition $Si_{2+x}Mn_{1-x}$ with diamond-like structure. The variant Si:15%Mn discussed here can be designated as $Si_{2.5}Mn_{0.5}$.

The HRTEM plane-view data of blocks along the planes of super-lattice modulation show two variants of substitution of silicon by manganese in the (110) monoatomic layers doped by these impurity. We assume as in [15] that negatively charged acceptor impurity atoms of manganese look darker in comparison with silicon atoms. One can see that in **Fig. 3a** (or **Fig. 2a** in [15]) Mn impurity ions are approximately half substituting silicon with intervals in length from one to several nm with continuous sequence of these ions. The second variant of the ordered substitution of silicon by manganese show **Fig. 3b** and **Fig. 5a**. Here in (110) Mn doped monoatomic layers the manganese atoms practically completely replace atoms of silicon. Presence in the set of blocks with sorts as in **Fig. 3a, b** and **Fig. 5a** of only two visible variants of substitution means that atoms of manganese in the layers doped by these impurity (110) settle down in the form of strips of an one-atom thickness, width from one to several nm, with a length stretched within all block. These manganese strips in the given (110) layer alternate with silicon strips of one-atom thickness and occupy about half of area of a layer (110) in the full consent with 15 % Mn according the spectral X-ray analysis and FMR data.

We are grateful to V.V. Podolskii and V.P. Lesnikov for preparation of samples, to D.A. Pavlov for his assistance in the experiments and helpful discussions. This work was supported by the Russian Foundation for Basic Research (project nos. 05_02_17362, 08_02_01222a, and 11_02_00855a), by the International Science and Technology Center (grant no. G1335), and by the Ministry of Science and Education of the Russian Federation (project nos. 2.1.1/2833 and 2.1.1/12029, program "Development of Research Potentialities in Higher Education Institutions," and contract no. 02.740.11.0672, federal program "Human Capital for Science and Education in Innovative Russia" for 2009–2013).